\documentclass[aps,prl,twocolumn,showpacs,superscriptaddress,floatfix,nofootinbib]{revtex4}

\usepackage{graphicx}
\usepackage{epsfig}
\usepackage[english]{babel}

\newcommand{\be}{\begin{equation}}
\newcommand{\bea}{\begin{eqnarray}}
\newcommand{\eea}{\end{eqnarray}}
\newcommand{\ee}{\end{equation}}

\begin{document}

\title{Adiabatic quantum computation and quantum phase transitions}

\author{Jos\'e Ignacio Latorre}
\author{Rom\'an Or\'us}
\affiliation{Dept. d'Estructura i Constituents de la Mat\` eria,
Univ. Barcelona, 08028, Barcelona, Spain.}
\date{\today}

\begin{abstract}

We analyze the ground state entanglement
in a quantum
adiabatic evolution algorithm designed to solve the
NP-complete  Exact Cover problem.
The entropy of entanglement
seems to obey  linear and universal scaling
at the   point where the energy gap becomes small,
suggesting
that the system passes near a quantum phase transition.
Such a large scaling of entanglement suggests that
the effective connectivity of the system diverges
as the number of qubits goes to infinity and that
this algorithm cannot
be efficiently simulated by classical means.
On the other hand, entanglement in Grover's algorithm is bounded
by a constant.

\end{abstract}
\pacs{03.67.-a, 03.65.Ud, 03.67.Hk}

\maketitle


Quantum adiabatic
computation \cite{farhi1} inherently brings the quantum system
near to a point where a quantum phase transition \cite{Sa99} takes place.
Entanglement is then expected to pervade the system as
we shall numerically give evidence
when quantum adiabatic computation is applied to the
Exact Cover problem, classically classified as  NP-complete.
Furthermore, scaling of the entropy of entanglement appears
to be linear for spin systems with non-local interactions,
sub-linear for the XX model, logarithmic for the quantum Ising
model \cite{kike1, kike2} and bounded for the Grover's adiabatic algorithm.
In order to substantiate these statements we first need
to recall three established results.


Large pure-state entanglement of the quantum register is a key element for
exponential speed-up of quantum computation. This result has been
made quantitative by Vidal  who has proven that a quantum
register, such that the maximum Schmidt number of any bipartition
is bounded, can be simulated efficiently by classical means. The
measure of entanglement proposed in ref. \cite{guifre} is $ E
\equiv \log_2 \chi $,  where $\chi$ is the maximum Schmidt number
of any bi-partitioning of the state. It can be further proved that
$E \ge S(\rho)$, where the von-Neumann entropy $S$ refers to the
reduced density matrix of any of the two sub-partitions. If a
$n$-qubit quantum register only uses little entanglement all along
the computation, that is $\chi={\rm poly}(n)$ at most, both the
quantum state as well as the action of the quantum gates on it can
be efficiently simulated by classical means. This implies that
exponential speed up is only possible if entanglement pervades the
quantum register at some point along the computation, that is, if
$\chi\sim \exp(n^a)$, with $a$ being a
positive constant, which is naturally satisfied if the entropy obeys
$S(\rho)\sim n^b$, for some positive constant $b$. Any algorithm designed 
to exponentially accelerate a classical computation must create exponentially large
$\chi$. An exponentially big $\chi$ is therefore a necessary, though not
sufficient, condition for quantum exponential speed-up.


On a second separate development, entanglement for the ground
state of many quantum spin chain systems has been proven to scale
at quantum phase transitions \cite{kike1, kike2} (see also
\cite{callan, preskill}). The entropy associated to tracing out
all but $L$ spins  out of an infinite spin chain displays
logarithmic scaling controlled by the central charges, $c$ and
$\bar c$, classifying the universality class of the phase
transition \cite{callan}: \be S_L={c+\bar c\over 6} \log_2 L\ .\ee
Results from field theory suggest that $d$-dimensional spin
systems should display a leading scaling behavior completely
determined by the area of the region separating the partitioning
of the system. For instance, when separating the system in the
interior and exterior of a sphere of radius $R$ and assuming an
ultraviolet cutoff $x_0$, the entropy of {\sl e.g.} the interior
is \be S=c_1 \left( R\over x_0\right)^{d-1} \ee where $c_1$
corresponds to a known heat-kernel coefficient \cite{kabat}. This
leading scaling behavior can be cast in terms of  the number of
spins in the system as \be S\sim n^{d-1\over d}\ .\ee Entanglement
only saturates for non-critical quantum systems in one dimension
\cite{kike2}.


The third element we need to introduce corresponds to the quantum
adiabatic computation framework introduced by Farhi et al. \cite{farhi1}.
The quantum register
is initially prepared on the ground state of a known initial
Hamiltonian $H_0$. The system is then made to evolve adiabatically
from this Hamiltonian to a new one $H_P$ whose ground state
codifies the solution to an {\sl e.g.} NP-complete problem
\be H(s(t))= (1-s(t)) H_0 + s(t) H_P \ . \ee
Slow evolution from $s(t=0)=0$  to $s(t=T)=1$ guarantees that
the system will not jump from the instantaneous ground state of the system to
the first excited state.
Quantum adiabatic computation is
efficient provided that the minimum gap along the adiabatic
evolution is only polynomially small in
the number of qubits.


It follows from the above arguments that quantum adiabatic
computation can be viewed as a time evolution in which there is
a flow along the parameter
space defining the Hamiltonian. At a given point $s_c$
the Hamiltonian approaches a quantum phase transition,
characterized by a vanishing energy gap. Exponential speed-up
needs large entanglement which is also expected at
some quantum phase transitions as discussed previously.
A quantum computer programmed to find the solution to
a given problem using adiabatic evolution does in fact
correspond to a system that passes near a quantum
phase transition. A quantum computation is thus equivalent
to the simulation of a very specific quantum phase transition.
Reversely, simulating a quantum phase transition is known
to be in general a hard problem that is in principle efficiently solved
by adiabatic evolution if the energy gap does not vanish exponentially
with the number of qubits.

We shall give support to the above picture by analyzing the
span of entanglement along a quantum adiabatic computation
applied to the Exact Cover problem, closely related to
the 3-SAT NP-complete problem. We shall indeed see that
entanglement seems to span over exponentially many states in
the computational basis and therefore the algorithm may be hard to simulate
in an efficient way using a classical computer.
This is a necessary (thought not sufficient)
condition for quantum exponential speed-up which
is apparently successfully verified in our case.

The NP-complete Exact Cover problem is a particular case of the
3-SAT problem and is defined as follows: given the $n$ boolean
variables $\{x_i\}_{i=1,\ldots n}$, $x_i = 0,1 \ \forall \ i$,
where $i$ is the bit index, we define a \emph{clause}
$C$ involving the three bits $i$, $j$ and $k$
 by the constraint $x_i + x_j + x_k = 1$. There are
only three assignments of the set of variables $\{x_i, x_j, x_k
\}$ that satisfy this equation, namely, $\{1,0,0\}$, $\{0,1,0\}$
and $\{0,0,1\}$.
An \emph{instance} of Exact Cover is a collection of
clauses which involves different groups of three qubits. The
problem is to find a string of bits $\{x_1, x_2 \ldots , x_n \}$
which satisfies all the clauses.

The Exact Cover problem can be mapped to finding the ground state of a
Hamiltonian $H_P$ in the following way \cite{farhi2}: given a clause $C$ define
the Hamiltonian associated to this clause as
\begin{eqnarray}
&&H_C=
\frac{1}{8}\left((1+\sigma_i^z)(1+\sigma_j^z)(1+\sigma_k^z)
 \nonumber \right.\\ &&+
(1-\sigma_i^z)(1-\sigma_j^z)(1-\sigma_k^z)
 +
(1-\sigma_i^z)(1-\sigma_j^z)(1+\sigma_k^z)
 \nonumber \\  &&+    \left.
(1-\sigma_i^z)(1+\sigma_j^z)(1-\sigma_k^z)
 +
(1+\sigma_i^z)(1-\sigma_j^z)(1-\sigma_k^z) \right)\nonumber \\
\end{eqnarray}
where $\sigma^z |0\rangle = |0\rangle$, $\sigma^z
|1\rangle = -|1\rangle$.
The quantum states of the
computational basis that are eigenstates of $H_C$ with zero
eigenvalue (ground states) are the ones that correspond to the bit
string which satisfies $C$, whereas the rest of the computational
states are penalized with an energy equal to one.
The problem Hamiltonian is constructed as the sum of all the
Hamiltonians corresponding to all the clauses in the
instance,
\begin{equation}
H_P = \sum_{C \ \in \ {\rm instance}} H_C \ . \label{hamil}
\end{equation}
The ground state of this Hamiltonian corresponds to the quantum
state whose bit string satisfies \emph{all} the clauses.
The original problem stated in terms of boolean logic has been cast into
the hard task of finding the ground state of a spin system
with non-local two and three body
interactions. The couplings depend on the particular chosen instance and, therefore, the spin
system has not an {\sl a priori} well defined dimensionality neither a
well defined lattice topology, in contrast with the usual
spin models (e.g. the anti-ferromagnetic Heisenberg model in a 2-dimensional square grid). 
This intrinsic relation between physical systems and
difficult computational problems is a well established fact. For example, the
ground-state search for some spin Hamiltonians defined on regular 
two-dimensional planar cubic lattices is proved to be an NP-hard problem
\cite{ba}. Notice that the systems considered in this paper, namely those
arising from the Exact Cover problem, differ a lot from those of \cite{ba}
as long as the detailed structure of the Hamiltonian is considered. Nevertheless, it is
a remarkable fact that the Exact Cover problem can be mapped to a ground-state
search of a spin-system, as in \cite{ba}, which provides some physical intuition.


Adiabatic evolution is carried by the s-dependent Hamiltonian
$H(s)$ as a linear interpolation between an initial Hamiltonian
$H_0$ and $H_P$: $H(s) = (1-s)H_0 + s H_P$, where the initial
Hamiltonian $H_0$ can be taken as a magnetic field in the $x$
direction
\begin{equation}
H_0 = \sum_{i = 1}^n \frac{d_i}{2}(1-\sigma_i^x) \ , \label{h0}
\end{equation}
where $d_i$ is the number of clauses in which qubit $i$ appears.
The ground state of
$H_0$ is an equal superposition of all the possible computational
states. Observe that $H(s)$ is, apart from a constant factor, a
sum of terms involving local magnetic fields in the $x$ and $z$
direction, together with two and three-body interaction coupling
terms in the $z$ component.

Our numerical analysis is based on the random generation of 300
instances for Exact Cover with only one possible satisfying
assignment for $n=6$ up to $n=20$ qubits. We produce the instances
by adding clauses at random until there is exactly one satisfying
assignment, starting over if we end up with no satisfying
assignments. According to \cite{farhi2}, these are believed to be
difficult instances for the adiabatic algorithm. For
every instance, we have constructed its corresponding
interpolating Hamiltonian and found the ground state for $s=0$ to
$s=1$ in steps of 0.01. We then consider a bipartition of the
system into two blocks of $n/2$ qubits
 and calculate the entanglement entropy between
the two blocks as a function of $s$. We have explicitly checked on
some instances that all possible partitions produce entropies of
the same order of magnitude (as expected from the non-locality of
the interactions) and chosen to work with the first $n/2$ versus
the rest.

\begin{figure}[h]
\centering
\includegraphics[angle=-90, width=0.49\textwidth]{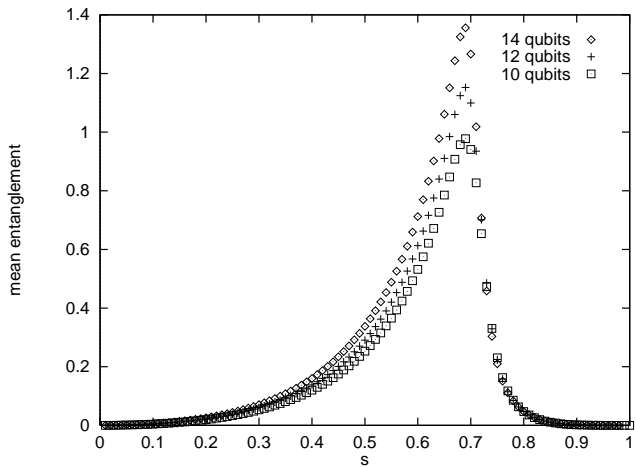}
\caption{Average over 300 instances of the entanglement entropy between two
blocks of size $n/2$  as a function
of the parameter $s$ controlling the adiabatic evolution. A
peak appears for $s_c \sim 0.7$. The plot also shows the increase of
the peak as the number of qubits grows $n=10,12,14$.
} \label{mix-300}
\end{figure}

The results we find for the scaling of
entanglement seem to agree with the idea that the system approaches
a quantum phase transition along its adiabatic evolution. For
each of the randomly generated Hamiltonians we observe a peak in
the entanglement entropy around a critical value of the parameter
$s_c \sim 0.7$. The average entropy shape over the 300
instances is represented in  Fig.
\ref{mix-300}.
\begin{figure}[h]
\centering
\includegraphics[angle=-90, width=0.49\textwidth]{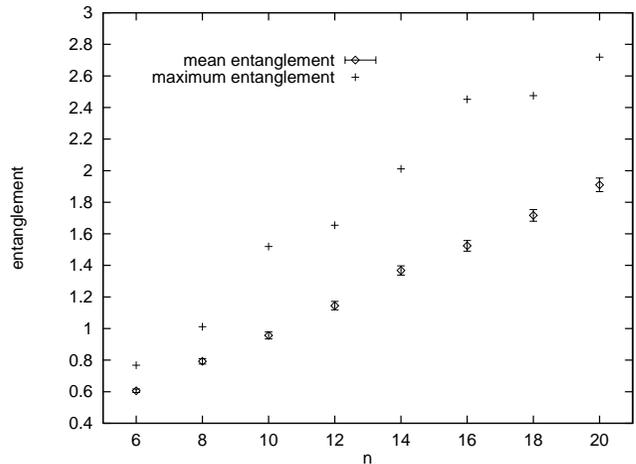}
\caption{Scaling of the entanglement entropy
 both for the worst case and for the average
over 300 randomly generated instances. Error bars give 95 per cent confidence
 intervals. The behavior appears to be
linear. } \label{ent}
\end{figure}

In order to analyze how entanglement scales at the critical value
$s_c$, we plot the maximum entropy of entanglement as a function
of the number of qubits, both for the worst case and for the
average over the 300 instances.

The numerical analysis shown in Fig. \ref{ent} apparently agrees with 
linear scaling and
matches the expectation that the Exact Cover problem can be
viewed as a spin system with highly non-local three-body couplings, and
therefore high effective dimensionality.
The points can be fitted by a function of the type $E(n)\sim .1\ n$.
This behavior would correspond to a nearest neighbor-like coupling
in $d\sim n$ dimensions, thus diverging as $n$ goes to infinity.
We note at this point that the evidence of large entanglement present in the
ground-state of the system does not say anything about the efficient
performance of the quantum adiabatic algorithm. Despite involving a highly
quantum-correlated system, the running time of the algorithm would still be
inefficient if the gap were exponentially small. This does not seem to be
the case, as we shall see, according to our simulations.

\begin{figure}
\centering
\includegraphics[angle=-90, width=0.49\textwidth]{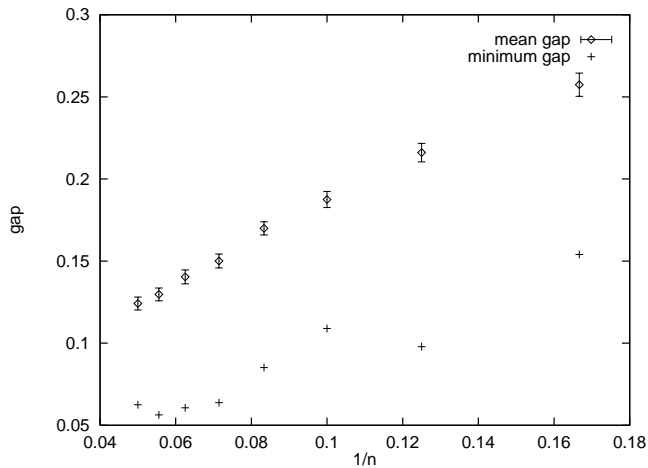}
\caption{Minimum gap versus the inverse size of the system, both
for the worst case and for the average over all  instances. Error bars give 95
per cent confidence intervals. The
behavior appears to be linear. } \label{gap}
\end{figure}

Our numerical analysis is also consistent with the work of Farhi
et al. in \cite{farhi2} where the minimum energy gap appears to
decrease as $g_{min}\sim 1/n$, as shown in Fig. \ref{gap}. It is
important to emphasize the difference between finding  scaling
laws for averages and analyzing the worst case. From the point of
view of characterizing a phase transition, averages over Exact
Cover instances follow quite well defined laws. The worst case is
harder to discuss as no systematic search of it can be done. The
worst case results can only be considered as consistent with the
polynomial vanishing of the energy gap. It is worth noticing that
the worst case, defined as the instance with a smaller minimum
gap,  brings also the higher entanglement as the system is passing
closer to the phase transition.
\begin{figure}
\centering
\includegraphics[angle=-90, width=0.49\textwidth]{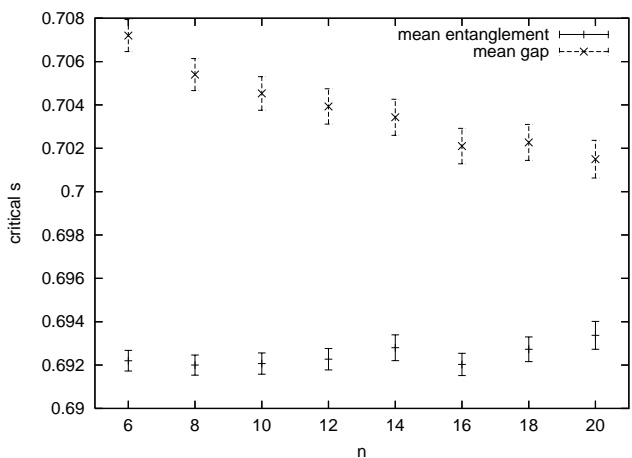}
\caption{Mean critical point $s_c$ for the minimum gap and for the
maximum entropy. Error bars give 95 per cent confidence intervals.
Both sets of points tend to approach as the size of the system is
increased.} \label{move}
\end{figure}
Moreover, the minimum gap takes place at the same place where
entanglement peaks. This phenomenon is illustrated in
Fig.\ref{move} where $s_c$ appears to converge to the same value
when $n\to\infty$ from above  for the minimum gap and from below
for the maximum entanglement.


The scaling of entanglement entropy as the system passes near a
quantum phase transition point is asymmetric, as is seen in
Fig. \ref{mix-300}. The growth of
entanglement is slower in the beginning of the evolution and fits
remarkably well a curve of the type $E\sim \log\vert \log
(s_c-s)\vert$, whereas the falling down of the peak is better
parametrized by a power law $E\sim \vert (s-s_c)\vert^{-\alpha}$ with
$\alpha\sim 2.3$. Both scaling behaviors improve as $n$ becomes
larger.


Grover's algorithm \cite{grover} does not produce exponential speed-up. It is
then arguable that entanglement, despite being necessary for having some
computational speed-up, should not play a relevant role
in this case. It is indeed possible to analytically address this
question. Let us cast Grover's algorithm into the adiabatic
evolution of the Hamiltonian
\begin{equation}
H(s) = (1-s)(I - |s\rangle \langle s|) + s (I-|x_0\rangle \langle
x_0|) \ , \label{grahm}
\end{equation}
where $|s\rangle \equiv \frac{1}{2^{n/2}} \sum_{x = 0}^{2^n-1}
|x\rangle$, $n$ is the number of qubits, and $|x_0\rangle$ is the
marked state \cite{roland, dam}. Other alternative definitions of the
Hamiltonian used in Grover's adiabatic algorithm (such as Hamiltonians
explicitely defined in terms of qubits) do not lead to significantly different
conclusions from the ones presented in this work. The computation takes the quantum
state from an equal superposition of all the possible
computational states directly to the state $|x_0\rangle$, as long
as the evolution remains adiabatic. The time the algorithm takes
to succeed depends dramatically on how we choose the
parametrization of $s$ in terms of time. The discussion on
entanglement can be reduced to analyze its dependence on $s$ since
the explicit dependence on time and its consequences (see
\cite{roland, dam} for further information about this topic) are
of no relevance.

It is straightforward to check that the Hamiltonian (\ref{grahm})
has its minimum gap between the ground and first excited states at
$s = 0.5$, which goes to zero exponentially fast as the number of
qubits in the system is increased. Therefore, this Hamiltonian
seems to undergo a quantum phase transition in the limit of
infinite size as $s=0.5$ becomes a non-analytical point (for more
on the Grover problem as a quantum phase transition, see
\cite{childs}). We consequently expect quantum correlations to be
maximum for this value of $s$. We present without proof
\cite{inprep} the result of the exact analytical calculation which
shows that, for any equally sized bipartition, the entanglement
entropy scales as
\begin{equation}
E(s=0.5, n\to \infty) = 1 - \frac{4}{\ln{2}} 2^{-n/2} \ ,
\label{scaling}
\end{equation}
so the entropy tends to $1$ for $s = 0.5$ as an square root in the
exponential of the size of the system, which is the typical factor
in Grover's quantum algorithm. Entanglement is also bounded for
all possible bi-partitions and no exponential speed up is present. We must 
remark that our analysis is based on the study of the
quantum state between succesive calls to the quantum oracle. Entanglement 
in the quantum register might become very high during the application of the
unitary black box, which will in turn depend on the specific searching
problem we wish to solve. This very general situation can not be adressed in detail
as it depends on the realization of the black box. As long as we restrict
ourselves to the situation between calls of the unitary oracle,
we see that entanglement is a bounded quantity (which is not necessarily the case along
the particular implementation of the quantum black box).

The main theoretical challenge in quantum computation theory
remains quantum algorithm design. It has been observed that
majorization theory seems to play an important role in the
efficiency of quantum algorithms \cite{latorre, orus1}.
Nevertheless, a relevant element for quantum computational
speedup seems to be entanglement \cite{ent1, ent2, ent4, guifre}.
Our results build on previous work \cite{farhi1, farhi2, kike1,
kike2, guifre} and suggest that entanglement grows exponentially
(as measured by $\chi$, the maximum rank of the reduced density matrices 
obtained over all possible bi-partitions) at a universal point along 
the adiabatic quantum evolution for the
Exact Cover problem, which limits the possibility of an efficient
classical simulation.
 At this point, the
system comes close to a quantum phase transition and
entanglement obeys
a scaling law that needs further investigation. For instance, it has also been
proved that entanglement in Shor's factoting algorithm diverges exponentially
fast in the number of qubits, which makes this algorithm difficult to simulate
classically as well \cite{inprep}.
Scaling of entanglement seems to be further related to the
effective connectivity of the system. Grover's problem
reduces to a two state problem and entanglement is bounded.
The quantum Ising, XX and Heisenberg spin chains show
logarithmic scaling. Higher dimensional spin models
obey faster scaling laws. The maximum connectivity
corresponds to non-local interactions (as those present
in the adiabatic evolution algorithm for the 3-SAT problem)
and entropy does approach its maximum scaling.
Computationally hard problems are thus associated to
quantum systems that present phase transitions where
entropy comes close to its maximum possible scaling.

\textbf{Acknowledgments:} We are grateful to J. Bergli, A.
Childs, M. A. Mart\'{\i}n-Delgado, E. Rico  and G. Vidal for
fruitful discussions. We acknowledge financial support from MCYT
FPA2001-3598, GC2001SGR-00065, IST-1999-11053, PB98-0685 and
BFM2000-1320-C02-01 and the  Benasque Center for Science.

\end{document}